\documentclass[
aip,
mre,
amsmath, amssymb,
reprint, 
twocolumn %%%%% Comment onecolumn to override one column format %%%%%%%
]{revtex4-1}

\usepackage{graphicx}% Include figure files
\usepackage{ulem}
\usepackage{dcolumn}% Align table columns on decimal point
\usepackage{bm}% bold math
\usepackage{xcolor}
\usepackage{braket} %量子算符宏包
\usepackage[utf8]{inputenc}
\usepackage[T1]{fontenc}
\usepackage{mathptmx}
\usepackage{etoolbox}
\usepackage{url}
\usepackage[colorlinks,linkcolor=blue,anchorcolor=blue,citecolor=blue]{hyperref}

\makeatletter
\def\@email#1#2{%
 \endgroup
 \patchcmd{\titleblock@produce}
  {\frontmatter@RRAPformat}
  {\frontmatter@RRAPformat{\produce@RRAP{*#1\href{mailto:#2}{#2}}}\frontmatter@RRAPformat}
  {}{}
}%
\makeatother
\begin{document}

%\preprint{AIP/123-QED}

%\title{Efficient machine learning for optimizing HHG intensity and ellipticity}

\title{Efficient optimization of plasma surface high harmonic generation by an improved Bayesian strategy}

% Force line breaks with \\
\author{Lili Fan}
\author{Ziwei Wang}
\author{Chenfei Liao}
\affiliation{School of Mathematics and Computer Science, Wuhan Polytechnic University, Wuhan 430023, China} 
\author{Jingwei Wang}
\email{wangjw@siom.ac.cn}
\affiliation{State Key Laboratory of High Field Laser Physics, Shanghai Institute of Optics and Fine Mechanics, Chinese Academy of Sciences, Shanghai 201800, China}
%\affiliation{China-Russian Belt and Road Joint Laboratory on Laser Science, Shanghai 201800, China}
%\affiliation{Collaborative Innovation Center of IFSA, Shanghai Jiao Tong University, Shanghai 200240, China}

\date{\today}

\begin{abstract}
Plasma surface high-order harmonics generation (SHHG) driven by intense laser pulses on plasma targets enables a high-quality extreme ultraviolet source with high pulse energy and outstanding spatiotemporal coherence. Optimizing the performance of SHHG is important for its applications in single-shot imaging and absorption spectroscopy. In this work, we demonstrate the optimization of laser-driven SHHG by an improved Bayesian strategy in conjunction with particle-in-cell simulations. A traditional Bayesian algorithm is first employed to optimize the SHHG intensity in a two-dimensional space of parameter. Then an improved Bayesian strategy, using the Latin hypercube sampling technique and a dynamic acquisition strategy, is developed to overcome the curse of dimensionality and the risk of local optima in a high-dimensional space optimization. The improved Bayesian optimization approach is efficient and robust in three-dimensionally optimizing the harmonic ellipticity, paving the way for the upcoming SHHG experiments with a considerable repetition rate. 

%Light carrying time-varying orbital angular momentum (OAM) is a recently discovered type of structured electromagnetic field, which is termed as a self-torqued field, compared with a typical vortex field whose OAM is static. While a self-torqued light is employed for manipulating the fast magnetic, topological, and quantum excitations, to increase its intensity and shorten the duration would be of great benefit. Here we demonstrate the generation of intense self-torqued harmonics and attosecond pulses in the relativistic regime, driven by two time-delayed relativistic vortex lasers with different OAMs $l_1$ and $l_2$. The OAM of the n-th harmonic spans $nl_1$ to $nl_2$, and the OAM of the attosecond pulses changes from $l_1$ to $l_2$. Such intense self-torqued harmonics and attosecond pulses may offer new possibilities in ultrafast spectroscopy.
\end{abstract}

\maketitle

\section{\label{sec:level1}INTRODUCTION}

Plasma surface high-order harmonics generation (SHHG) is an extreme nonlinear process taking place on the plasma surface when irradiating an intense laser pulse to a plasma target.\cite{RevModPhys.81.445,dromey_high_2006,PhysRevLett.96.125004} The dense electrons on the plasma surface will efficiently reflect the laser field, which is called a plasma mirror.\cite{thaury_plasma_2007} Since the intensity of the driving laser is relativistic ($I>1.38\times10^{18}$ W/cm$^2$ for a 800~nm laser), the electron sheet will oscillate along the normal direction of the surface with a speed close to the light speed $c$ in vacuum. This gives rise to the emission of extreme ultraviolet (EUV) or soft x-ray photons in the specular direction, due to the relativistic Doppler effect.\cite{einstein_zur_1905,bulanov_interaction_1994,lichters_short-pulse_1996} As these emissions occur periodically, their spectrum exhibits a high harmonic structure, similar to the gas high harmonic generation.\cite{science.1059413,hentschel_attosecond_2001,science.1132838} SHHG is characterized by a high conversion efficiency and excellent spatiotemporal coherence.\cite{RevModPhys.81.445} Therefore, SHHG provides great potential in single-shot coherent diffraction imaging,\cite{PhysRevLett.99.098103,PhysRevLett.103.028104} x-ray absorption spectroscopy,\cite{hutten_ultrafast_2018} and generating intense attosecond pulses.\cite{Xu20,yeung_experimental_2017,PhysRevE.102.061201} 

There is no doubt that optimizing the performance of SHHG is beneficial for those important applications. For instance, single-shot coherent diffraction imaging in nonrepetitive experiments requires a high flux of coherent harmonics,\cite{PhysRevLett.103.028104}
and harmonics with controllable polarization are particularly useful for controlling magnetization dynamics and probing chiral molecules.\cite{boeglin_distinguishing_2010,PhysRevLett.122.157202,PhysRevLett.118.013002} However, optimizing SHHG is now still challenging because SHHG is a typical multi-parameter process. One needs to carefully balance the parameters in an $N$-dimensional space, including laser incidence angle, inclination angle, contrast ratio, pre-plasma length, plasma density, etc. A full $N$-dimensional scanning of parameters is too expensive in terms of time-consuming and experiment source, for both numerical simulations and real experiments. 

%for and  The usual approach to optimization is to perform a series of single variable (one-dimensional, 1D) scans in the neigh-bourhood of the expected optimal settings.

%Although a few physical models can in principle provide a prediction of the scaling law, there is still a lack of an efficient way to give accurate parameters for customized requirements. 

Recently, the Machine learning (ML) technique has been imported into the field of laser-plasma physics to address such challenges. ML is a powerful technique that can learn from known data and generalize to unseen data. It first trains a mathematical model based on a given dataset and then makes predictions or classifications on new data. After limited iterations, the model becomes "intelligent" enough to provide the optimal parameters.\cite{10.1126/science.aaa8415, YANG2020295} Since it was developed, ML quickly become an important and helpful tool in a variety of branches of physics. Because of its huge impact on physics, the 2024 Nobel Prize in Physics was awarded to ML with artificial neural networks.\cite{nobel2024}
%and it is not strange to see it won the 2024 Nobel Prize in Physics %2024 was awarded to John J. Hopfield and Geoffrey E. Hinton for their contributions to ML with artificial neural networks.
%2024 Nobel Prize in Physics was awarded to machine learning (ML) techniques are fast developed and hot, 
%ML has been also imported into the field of laser-plasma physics to address the challenges of multi-parameter optimization~[]. 
In the field of laser-plasma interactions,  ML has shown a great ability to control and optimize the process of multi-parameters.  
%Since being imported, ML has shown a great ability to control and optimize the process of laser-plasma interactions. 
For example, ML approaches based on genetic algorithms have been employed to optimize the wavefront\cite{he_coherent_2015} 
or temporal shape\cite{PhysRevAccelBeams.22.041303} 
of the driving laser in a laser wakefield electron acceleration experiment. Neural-network-based ML approaches were successfully applied in laser-driven ion acceleration.\cite{10.1063/5.0045449} 
%Compared with genetic algorithms and neural networks, Bayesian optimization is more popular since it requires less data to train the model. 
ML technique of Bayesian optimization has been used in laser-driven electron acceleration experiments\cite{shalloo_automation_2020,PhysRevLett.126.104801} 
and laser-driven ion acceleration experiments.\cite{PhysRevResearch.3.043140}
However, studies on optimizing SHHG employing ML techniques have not yet been reported.

In this work, we utilize an improved Bayesian optimization (BO) algorithm in conjunction with particle-in-cell (PIC) simulations to optimize the performance of SHHG. We first report the optimizing harmonic intensity in a two-dimensional space using a traditional BO algorithm. Then an improved BO approach, using the Latin hypercube sampling technique and a dynamic acquisition strategy, is introduced to solve the problem of the curse of dimensionality and the risk of local optima in a high-dimensional space optimization. The improved BO approach is efficient and robust in multi-dimensionally optimizing the harmonic ellipticity.  

\section{Bayesian optimization of harmonic intensity}

Bayesian optimization is an advanced ML technique that is particularly suited to optimize functions with forms of unknown structure.\cite{7352306,PhysRevAccelBeams.27.084801} It works as follows. BO treats the objective function as a random function and places a prior over it. The prior captures beliefs about the behavior of the function. After gathering the function evaluations, which are treated as data, the prior is updated to form the posterior distribution over the objective function. The posterior distribution, in turn, is used to construct an acquisition function that determines the next query point.
The core in BO is how to define the prior/posterior distribution over the objective function. Gaussian process regression (GPR) is one of the most common methods to do this job.\cite{doi:10.1142/S0129065704001899}
%surrogate model because of its flexibility and ability to provide both mean predictions and uncertainty estimates. 
%The core component of Bayesian optimization is the construction of a surrogate model and an acquisition function. It approximates the objective function by building a prior probability model and uses the acquisition function to balance the trade-off between exploration and exploitation.
%The BO process begins with the selection of a surrogate model. 

\begin{figure}
\centering
\begin{center}
\includegraphics[width=0.99\linewidth]{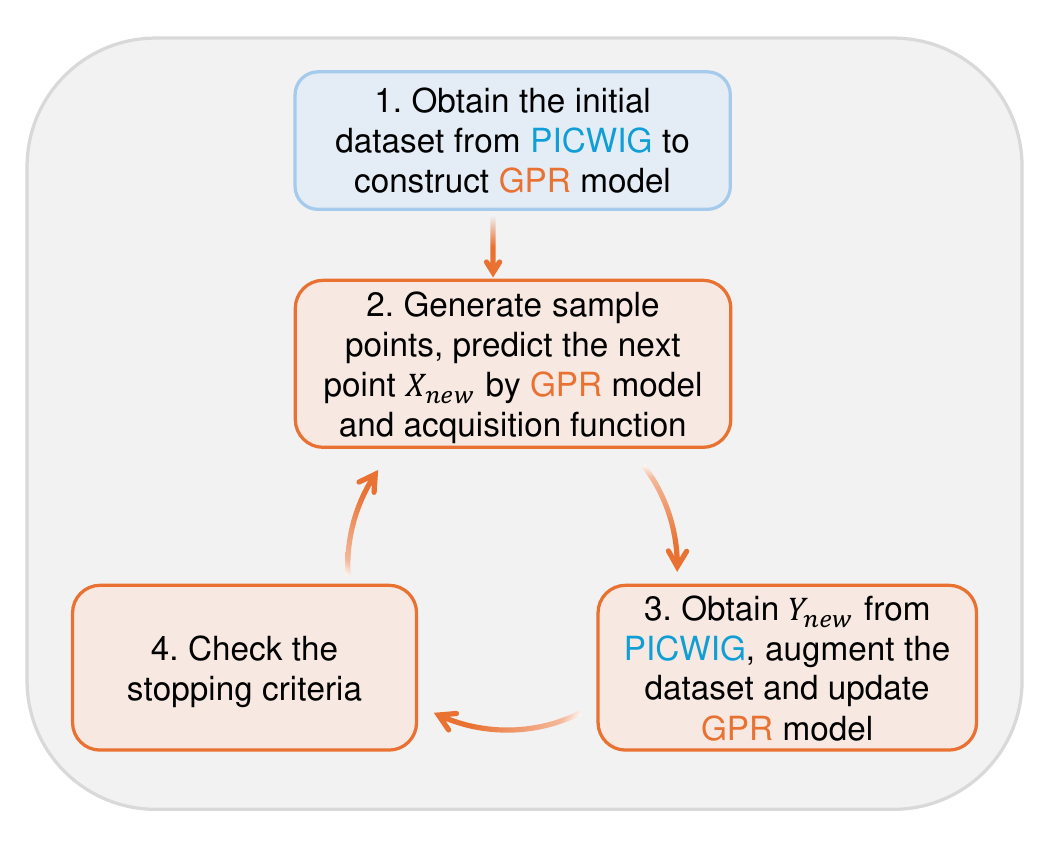}
\caption{The workflow of Bayesian optimization based on Gaussian process regression.}
\label{fig:flowchart}
\end{center}
\end{figure}

Here we developed an integrated GPR-based Bayesian optimization code, employing the tools from the open-source Scikit-Learn library,\cite{scikit-learn} to optimize the intensity and ellipticity of the plasma harmonics. The objective function values are obtained from the 1D PIC code PIGWIG\cite{Rykovanov_2008} in real-time. An efficient interface between the BO code and PIGWIG is established. The flowchart of the integrated code is presented in Fig.~\ref{fig:flowchart}. First, $N$ initial points $X$ are selected using a uniform sampling method and the corresponding objective function values $Y$ are obtained from the PIC simulations, which together serve as the given dataset to construct the GPR model. The generated samples in the space are then fed into the GPR model to predict $X_{new}$, which maximizes the acquisition function. $X_{new}$ is later on transferred to PIGWIG to obtain the new objective function value $Y_{new}$, which is then returned to the BO code. The GPR model is updated based on the new dataset, which includes $\{X_{new}$, $Y_{new}\}$. Steps 2 and 3 will repeat until the convergence criteria is met. At this stage, the BO algorithm is a traditional one.

%Initially, a prior distribution is assumed over the function space. 
%As the objective function is evaluated, this prior is updated to a posterior distribution that better approximates the true function. 
%In this iterative process, the GPR model is updated with new data.  

%A GPR is defined by a mean 
%function and a covariance function. 

%A comprehensive flowchart of the GRP-based Bayesian optimization is shown in Fig.~\ref{fig:flowchart}. Here each objective function value $Y$ is obtained in real-time from the particle-in-cell simulations. 

%In the parameter optimization process, grid search is frequently utilized to maximize two variables. 
%Grid search is a systematic search method whereby each variable is traversed in its entirety across the entire parameter space, 

We perform the traditional BO to optimize harmonic intensity in a two-dimensional space. The high-order harmonics are generated by obliquely irradiating a $p$-polarized intense laser pulse onto the surface of a plasma target. In the 1D PIC simulations, the laser pulse duration is $\tau=6\rm{T_L}$ and its wavelength is $\rm{\lambda_L}=800~\rm{nm}$, with $\rm{T_L}=2.67~\rm{fs}$ the laser period. The laser peak intensity is $5.4\times 10^{19}~\rm{W}/\rm{cm}^2$ (the corresponding normalized amplitude $a_0=5$). The distribution of the plasma density is shaped with a linear function $n_e = n_0x/L_{\rm{pre}}$ for $x<L_{\rm{pre}}$ and then keeps constant as $n_e=n_0=100\rm{n_c}$.  Here $x$ is the space coordinate and $n_c\approx1.7\times10^{21}/\rm{cm}^3$ is the plasma critical density for the 800~nm laser. The grid size is $0.001\rm{\lambda_L}$. A virtual detector is placed in the specular reflection direction to collect the reflected laser fields. The harmonic spectra are obtained by Fourier transforming of the reflected laser fields. It should be noted that the frequency of a specific harmonic order changes slightly for different parameters of laser or plasma. This brings non-negligible errors when optimizing the intensity of a specified order of harmonic. We thus use the intensity of an attosecond pulse, which is synthesized from a band (10~$\sim$~20$\rm{\omega_L}$) of the harmonics, as the target function of optimization. Here $\rm{\omega_L}=2\pi c/\lambda_L$ is the laser frequency. In this section two parameters, laser incident angle $\theta$ and pre-plasma length $L_{\rm{pre}}$, are the free variables for optimizing. 

\begin{figure}
\centering
\begin{center}
\includegraphics[width=0.99\linewidth]{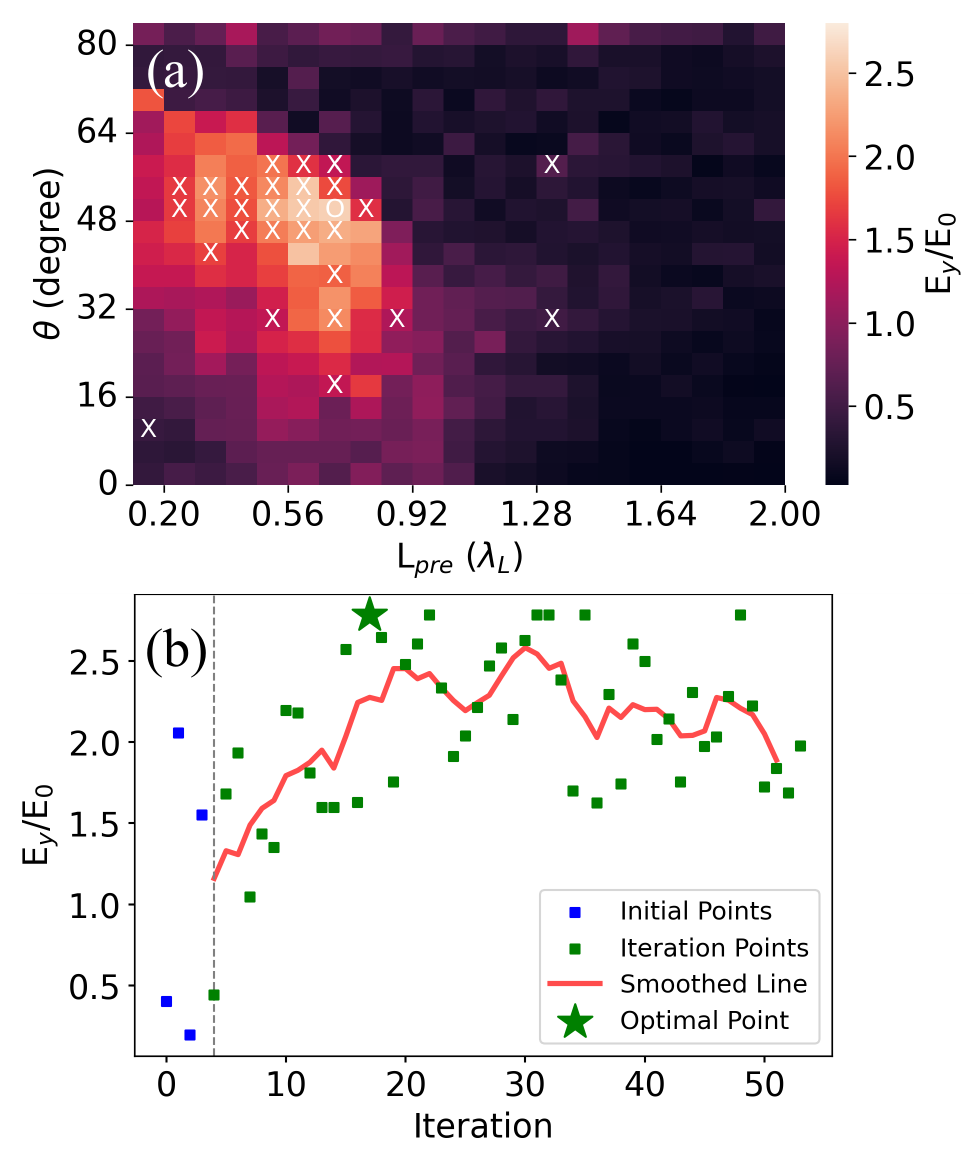}
\caption{(a) Distribution of the electric field E$_y$ of the attosecond pulse in the 2D parameter space (incident angle $\theta$ and preplasma length $L_{\rm{pre}}$). The data is obtained using a grid search method in conjunction with PIC simulations. The '$\times$' symbols mark the predicting points during the BO routine and the optimal condition for maximizing intensity is marked with 'O'. (b) Dependence of electric field E$_y$ of the attosecond pulse on the iteration number using BO. The blue points indicate the initial dataset and the green ones are the iteration points. The red smoothed line is plotted using the average of every 5 neighbor iteration points. The optimal point is marked with a star.}
\label{fig:BO}
\end{center}
\end{figure}

Before using Bayesian optimization, a reference dataset of 400 PIC simulations is produced using the grid search method.  Grid search is a conventional optimization way of scanning systematic parameters with a fixed increment. We select 20 uniform points in the $\theta$ space (0~$\sim$~$\pi/2$) and 20 uniform points in the $L_{\rm{pre}}$ space (0.1$\rm{\lambda_L}$$\sim$~2.0$\rm{\lambda_L}$) as the input grids for the PIC simulations. Figure~\ref{fig:BO}(a) shows the 2D dependence of the attosecond pulse intensity on the incident angle and preplasma length. One can see that the intensity maximizes at the grid point ($\theta=48$ degree, $L_{\rm{pre}}=0.7\lambda_L$). 

\begin{figure}
\centering
\begin{center}
\includegraphics[width=0.99\linewidth]{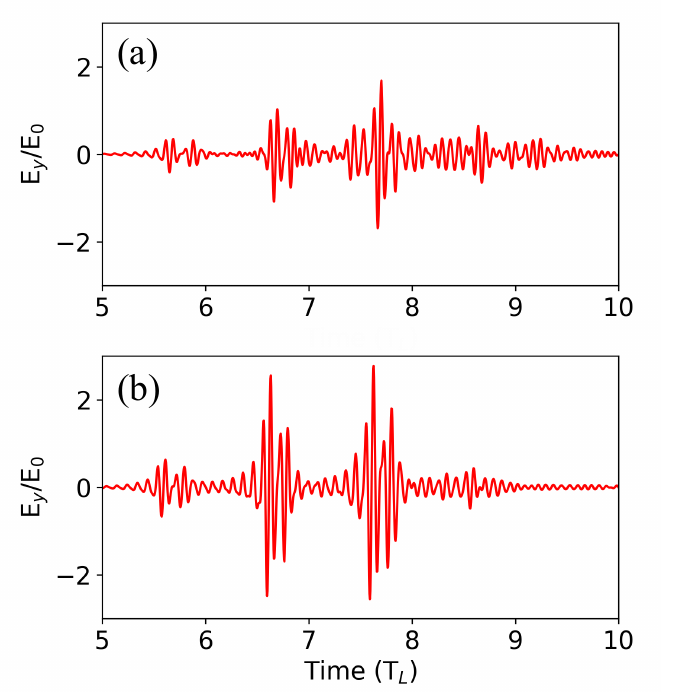}
\caption{The electric field E$_y$ of the attosecond pulses for different parameters (a) $\theta=38.75$ degree, L$_{\rm{pre}} = 0.423~\lambda_L$; 
%(b) $\theta=43.25$ degree, L$_{\rm{pre}} = 0.575~\lambda_L$; 
(b) $\theta=50.05$ degree, L$_{\rm{pre}} = 0.651~\lambda_L$. The latter case is the optimized one predicted by BO.}
\label{fig:intensity}
\end{center}
\end{figure}

Subsequently, the intensity optimization is conducted utilizing the BO method. The input parameters are first standardized to ensure that they work within the same scale range. Four initial points are selected using a uniform sampling method which ensures the reproducibility of the optimization. Figure~\ref{fig:BO}(b) shows the optimizing process. The 4 initial points are colored in blue. One can see that after 15 iterations the BO code reaches the optimized parameters ($\theta$=50.05 degree, $L_{\rm{pre}}$=0.651$\lambda_L$, labeled with a green star in the figure) and the normalized intensity of the attosecond pulse is 2.78, compared with 2.60 after scanning 400 points using the grid search method. The iteration points during the optimization routine are sampled on the full search grid in Fig.~\ref{fig:BO}(a). The optimal condition for maximizing intensity is marked with 'O'. 

The shape of the attosecond pulse generated from the optimal parameter is presented in Fig.~\ref{fig:intensity}(b). The duration of the attosecond pulse is about $200~\rm{as}$ and the normalized intensity is $2.78$, compared with an unoptimized case in Fig.~\ref{fig:intensity}(a), in which the normalized intensity of the attosecond pulse is around $1.6$. The peak of the electric field $\rm{E_y}$ of the attosecond pulse is the objective function in the optimization. $\rm{E_0}=m_e c \omega_L/e$ is the normalized unit, with $m_e$ the electron mass and $-e$ the electron charge. 
%The efficiency from the laser energy to the attosecond pulse energy is about $10^{-5}$.

Although Fig.~\ref{fig:BO}(b) confirms that BO can find the optimal parameters quickly, it will not keep stable around the optimal parameters afterward but oscillates with a big amplitude as the red line shows. It means that the traditional BO does not change its strategy when the best result so far has been found. This is mainly because of the use of a static acquisition function. Such an optimization, however, is not a favored manner for a real-time experiment, which wants to stay around the optimal conditions after the optimization. To overcome this defect, an improved BO strategy has been developed.

\section{Improved Bayesian strategy to optimize harmonic ellipticity}

The real laser-plasma experiment is more likely a multi-parameter process. Optimizing such a process is challenging for the above classical BO algorithm, in terms of two issues. The first one is the so-called curse of dimensionality in grid sampling. In a low-dimensional space, grid selection of sample points is an effective method that allows for a systematic exploration of all possible combinations within the parameter space. However, if this method is utilized, the size of the kernel matrix in the GPR model grows quadratically with the amount of data, leading to exponential expansion in high-dimensional scenarios.\cite{10.1145/3545611} This brings an expensive load on the computing source. The second issue is the limitation of the static acquisition functions in high dimensionality. The acquisition function is used to select the next experimental point in each iteration. In a high-dimensional space, the fixed hyperparameters in the static acquisition functions will impede the effective balancing of exploration and exploitation, thereby hindering the ability to adapt to the varying stages of the optimization process. Specifically, the optimization process either misses exploration opportunities due to being overly conservative or struggles to maintain stability due to being overly aggressive.
%Specifically, it is too conservative at the early stage of optimization and becomes overly aggressive at the later stage, thus putting BO in the trouble of local optima.
%Specifically, it is too conservative at the early stage of optimization, resulting in insufficient exploration and the inability to identify optimal regions within the parameter space. On the other hand, it becomes overly aggressive at the later stage of optimization, by continuing to explore unknown areas while neglecting the possibility of finding superior solutions in regions that have already been explored.

To avoid the curse of dimensionality in grid sampling, the Latin hypercube sampling (LHS) technique is employed.\cite{doi:10.1080/00401706.2000.10485979} LHS first divides each dimension into several intervals and then randomly selects a point within each interval. In such an algorithm, the combinations of all dimensions are independent and the coverage of the parameter space is more uniform. Fewer but more representative sample points are generated by LHS, which effectively mitigates the kernel matrix expansion.

To solve the limitation of the static acquisition functions, a dynamic acquisition strategy (DAS) is introduced into the present BO code. The DAS combines multiple acquisition functions, each of which generates its own suggested experimental point during the iterations. The algorithm assesses the performance of each acquisition function based on the updated model and adjusts its probability in the subsequent sampling event. Consequently, the better-performing acquisition function is given a higher selection probability and dominates in the subsequent sampling process. Moreover, DAS 
will dynamically modify the hyperparameters of each acquisition function, which effectively reduces the risk of falling into local optima.
%These functions are initiated with fixed hyperparameter values, which control the initial behavior of the search process. As experiments progress, the values of hyperparameters are gradually adjusted, shifting the behavior of the acquisition functions from exploration to exploitation. 
%This dynamic adjustment mechanism not only significantly speeds up optimization but also effectively reduces the risk of falling into local optima, thereby enhancing overall stability.

\begin{figure}
\centering
\begin{center}
\includegraphics[width=0.99\linewidth]{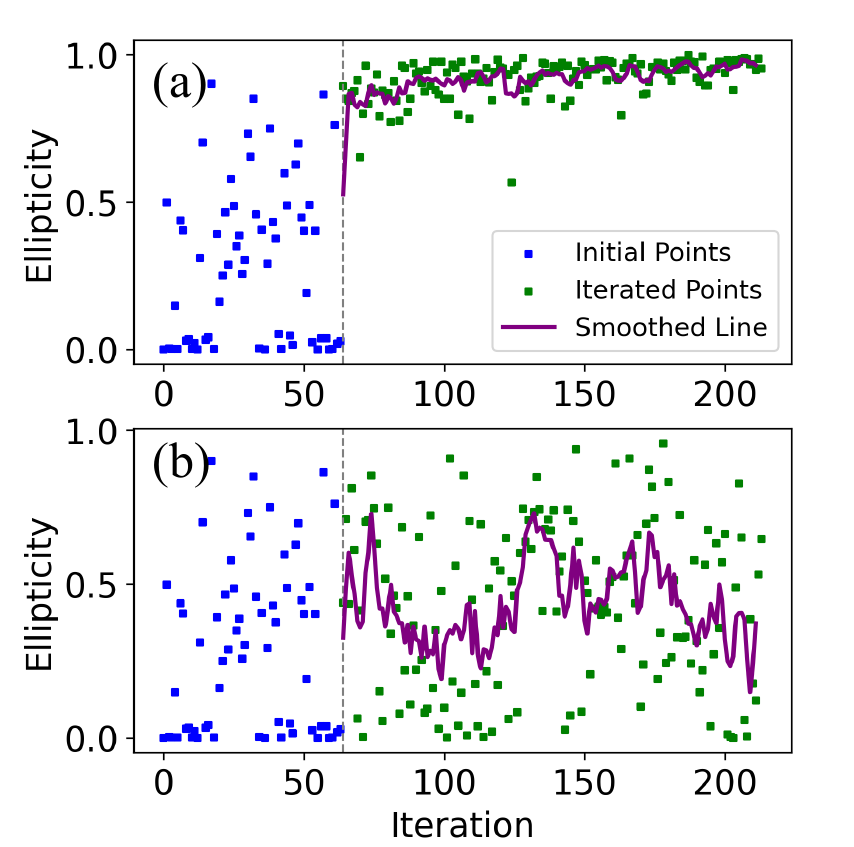}
\caption{Dependence of the ellipticity of the attosecond pulse on the iteration number using (a) an improved DAS-BO and (b) a classical BO. The blue points indicate the initial dataset and the green ones are the iteration points. The purple smoothed line is plotted using the average of every 5 neighbor iteration points.}
\label{fig4}
\end{center}
\end{figure}

The developed DAS-BO code is employed to optimize the ellipticity of the harmonics. Elliptically polarized harmonics could be generated when irradiating a linearly polarized laser pulse on the plasma surface with an incident angle $\theta$ and inclination angle $\alpha$.\cite{10.1063/5.0057689} The inclination angle is defined as the angle between the incidence plane and the polarization plane. The preplasma length also has an impact on the ellipticity of the harmonics.\cite{WOS:000381771900001,zhou2024}. Therefore, the harmonics ellipticity is optimized in a three-dimensional space $\{$ incidence angle $\theta$, inclination angle $\alpha$, preplasma length $L_{\rm{pre}}$  $\}$. Usually, the ellipticity of the harmonics or attosecond pulse is low as the driven laser is linearly polarized. The parameters $\{$$\theta$,  $\alpha$,  $L_{\rm{pre}}$$\}$ should be well-matched thus a high ellipticity could be achieved.

Here we use the ellipticity $\epsilon$ of an attosecond pulse as the target function of optimization. The ellipticity of the attosecond pulse is calculated using the Stokes parameters of the electric field of the attosecond pulse.\cite{wang_intense_2019} $\epsilon=0$ means linear polarization and $\epsilon=1$ means circular polarization. A high ellipticity is what we pursue during the optimization. Figure~\ref{fig4}(a) shows the dependence of ellipticity on the iteration number, while 64 initial points are used here. The DAS-BO quickly gets close to the optimal parameters after a few iterations and then almost keeps stable. The soothed line, which is plotted using the average of every 5 neighbor iteration points, indicates the iterating stability. In contrast, the process of a traditional Bayesian optimization has considerable fluctuations, without finally achieving stability, as shown by Fig.~\ref{fig4}(b). Therefore, the dynamic acquisition strategy helps BO to realize a stable iterating.

\begin{figure}[t]
\centering
\begin{center}
\includegraphics[width=0.99\linewidth]{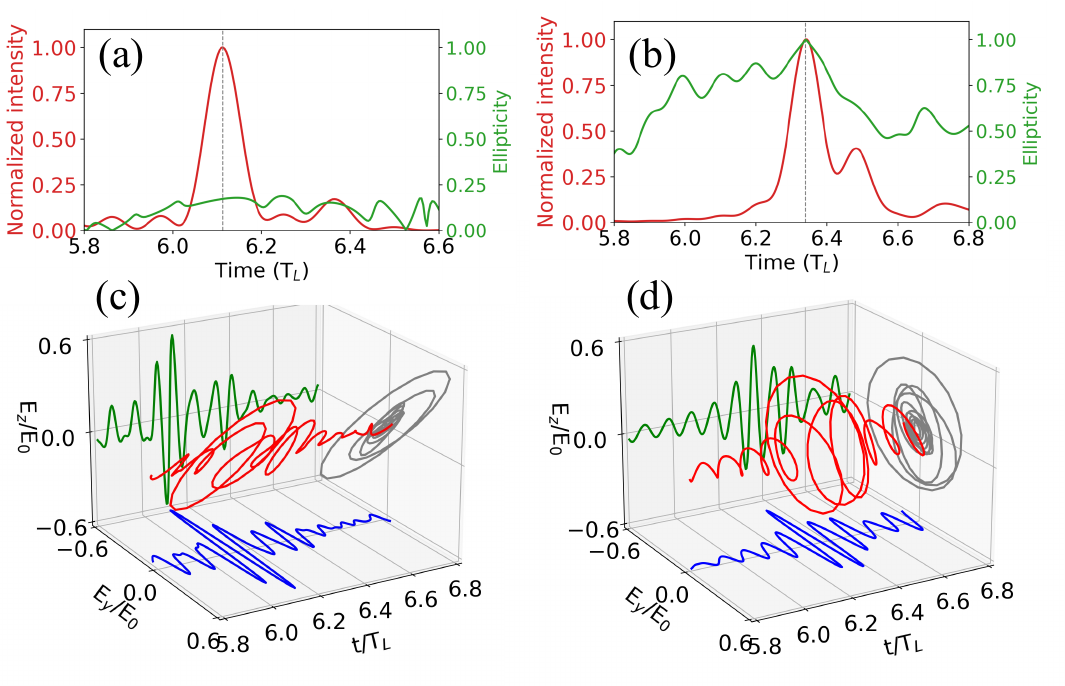}
\caption{(a) The ellipticity and (c) the structure of the attosecond pulse for a case of low ellipticity $\epsilon=0.17$ using the parameters: incident angle $\theta=7.01$ degree, inclination angle $\alpha=46.60$ degree, L$_{\rm{pre}} = 0.72~\lambda_L$. (b) and (d) are the same plots but for a case of high ellipticity $\epsilon=0.98$ using the parameters: incident angle $\theta=46.27$ degree, inclination angle $\alpha=50.68$ degree, L$_{\rm{pre}} = 0.76~\lambda_L$. In (a) and (b), the red line is for the normalized intensity of the attosecond pulse while the green line is for the ellipticity. Since the ellipticity changes with time, we use its value when the intensity peaks.}
\label{fig5}
\end{center}
\end{figure}

Figure 5 presents the ellipticity of the pulse and the corresponding structure of the electric field. Figure~\ref{fig5}(a) and (c) are for the case of a low ellipticity with $\epsilon=0.17$. The attosecond pulse is almost linear polarized like the incident laser, but with a rotation of the polarization plane. A case of high ellipticity with $\epsilon=0.98$ is demonstrated in Fig.~\ref{fig5}(b) and (d). The attosecond pulse is almost circularly polarized, although the driven laser is linearly polarized. Such a conversion can be understood like this. When one rotates the polarization plane of the incident laser, its electric field can be decomposed into $s$ and $p$ components. Each of them will generate their own harmonics. Changing the preplasma length will affect the phase difference between the two components' harmonics~\cite{zhou2024} and changing the inclination angle will affect the relative intensity of the two components' harmonics.\cite{10.1063/5.0057689}. When the preplasma length and the inclination angle are well matched, it is possible to generate a circularly polarized attosecond pulse using a linearly polarized driven laser. 
%The present study shows how to use the ML technique to achieve the matched condition.

%The advanced DAS-BO code is employed to optimize the ellipticity of the harmonics. Elliptically polarized harmonics could be generated when irradiating a linearly polarized laser pulse on the plasma surface with an incident angle $\theta$ and tilted angle $\alpha$~[]. The tilted angle is defined as the angle between the incidence plane and the polarization plane. The preplasma length also has an impact on the ellipticity of the harmonics, as explained by our previous work~[]. Therefore, the harmonics ellipticity is optimized in a three-dimensional space $\{$ incidence angle $\theta$, tilted angle $\alpha$, preplasma length $L_{\rm{pre}}$  $\}$. Here we also use the ellipticity of an attosecond pulse as the target function, instead of a specific harmonic order.

%The experimental results demonstrate that the DAS-BO performs exceptionally 
%well in optimizing HHG ellipticity. It significantly enhances the optimization effect with fewer iterations and 
%gradually stabilizes in subsequent iterations. 
%This comparison validates the superiority of the improved Bayesian optimization algorithm in exploring the parameter space, 
%enabling faster and more accurate identification of the optimal solution.  

\begin{figure*}
\centering
\begin{center}
\includegraphics[width=0.8\linewidth]{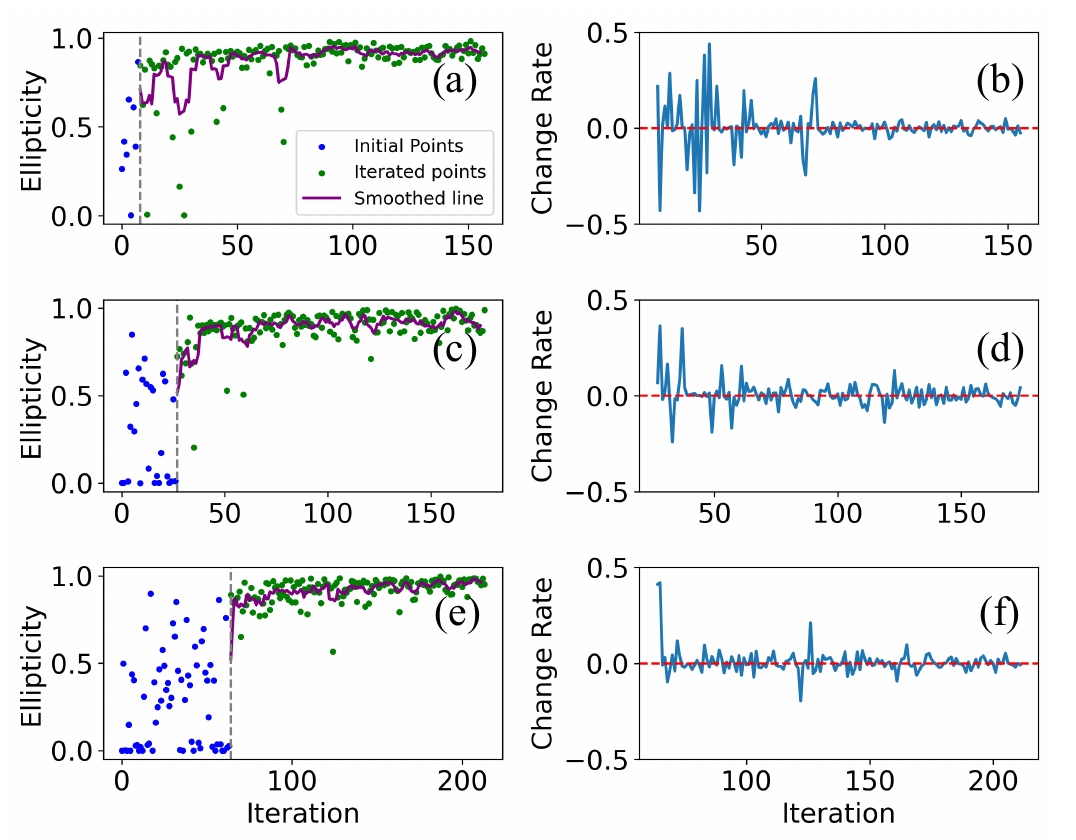}
\caption{The different BO processes (left column) and change rates of the ellipticity between two iterations (right column), for different numbers of initial points. The numbers of initial points are 8 for (a) and (b), 27 for (c) and (d), and 64 for (e) and (f). The change rate indicates how fast the ellipticities of the neighbor iteration points change. A smaller change rate means the BO process is more stable.}
\label{DASBO}
\end{center}
\end{figure*}

\section{DISCUSSION AND CONCLUSION}
Now we check the performance of DAS-BO on different numbers of initial points. Figure~\ref{DASBO} shows the iteration process and the change rate of the ellipticity between two iterations, for different numbers of initial points. The change rate indicates how fast the ellipticities of the neighbor iteration points change. A smaller change rate means the BO process is more stable. 
%One can see that the more initial points, the faster the optimization becomes stable.  
It can be seen that as the number of initial points increases, the optimization becomes stable faster. %This result highlights the critical impact of the number of initial points on optimization performance. 
More initial points enable the algorithm to cover the parameter space more comprehensively at the outset, thereby providing more information for the subsequent exploration. Therefore, the model can then effectively identify potential optimal regions at the early stage and reduce the uncertainty in the later iterations. However, in SHHG experiments, choosing the number of initial points is a key trade-off that should be carefully balanced, since the initial points also cost. On the other hand, experience and prior knowledge can be utilized to determine a reasonable number of initial points. 

In conclusion, an advanced DAS-BO strategy is developed to optimize the performance of SHHG. It is efficient and robust in three-dimensionally optimizing the harmonic intensity and ellipticity. The code will be further integrated into the SHHG experiment control platform.  Considering the 1Hz repetition rate of a 200TW laser system,\cite{WU2020106453} the next generation of SHHG experiments can achieve a good performance in a few minutes, paving the way for the practical use of SHHG.

\section*{DATA AVAILABILITY}

The data supporting this study's findings are available from the corresponding author upon reasonable request.

\section*{ACKNOWLEDGMENTS}
This work was supported by the National Key Research and Development Program of China (Grant No.~2023YFA1406804), the National Natural Science Foundation of China (No.~11991074 and 11871388), the International Partnership Program of the Chinese Academy of Sciences (Grant No.~111GJHZ2023060GC), and the Strategic Priority Research Program of the Chinese Academy of Sciences (Grant No.~XDA25051100, XDA25010100, XDB0890304).

\section*{REFERENCES}
\nocite{*}
\bibliography{aipsamp}% Produces the bibliography via BibTeX.
% ****** End of file aipsamp.tex ******
\end{document}